# ADOPTION OF FREE AND OPEN SOURCE SOFTWARE USING ALTERNATIVE EDUCATIONAL FRAMEWORK IN COLLEGE OF APPLIED SCIENCES

Mrs. Raya Al-Hajri[*1], Mrs. Ghada Al-Mukhaini[*2], Mr. Rajasekar Ramalingam [*3]

*Abstract* - The adoption of Free and Open Source Software (FOSS) in educational institutions is increasing day by day. Many countries are insisting the use of FOSS in their government sectors and few are in the process of adopting FOSS strategies. The reasons for adopting FOSS are: total cost ownership, free to make copies and distribution, software legality, reliability, availability, performance, security and other pedagogical and administrative benefits. In this research paper, the alternative educational framework has been used to identify the FOSS tools and software by considering the various strategic areas, polices and phased approach to identify the suitable FOSS tools for the courses of various specializations of Information Technology programme in College of Applied Sciences, Sultanate of Oman. As a result of this research, a list of alternative FOSS tools and software for the courses of IT programme has been identified and recommended.

## I. INTRODUCTION

The Ministry of Higher Education (MoHE) [1] offers, Bachelor of Information Technology (BIT) degree, a four year degree programme, through the college of applied sciences (CAS) in various regions of Sultanate. The information technology (IT) programme of CAS offers four different specializations namely: Software Development, Data Management, IT Security and Networking [2]. The courses in all the specializations are designed with a considerable percentage of practical and training works, which are carried out by using the proprietary software. CAS uses different modern technologies to distribute knowledge to students; however, most of these technological tools and software are vendor based [5], which makes the students more dependent on the institution for software installation and up gradation. CAS also needs to plan for the budget to purchase the necessary licenses for the students due to the payment of fee for commercial license agreement and still the buyer do not receive the rights to copy, modify, or redistribute the software without royalty or additional cost [3].

* Faculty, Department of Information Technology, College of Applied Sciences – Sur, Sultanate of Oman. [1][raya.sur@cas.edu.om]
[2][ghada.sur@cas.edu.om] [3][rajasekar.sur@cas.edu.om]

Based on the agreement of WTO with the Sultanate of Oman, there are restrictions for using the proprietary software even for educational purpose without license [4].

MITRE report [6] defines FOSS as; "the word free in FOSS refers not to fiscal cost, but to the autonomy rights that FOSS grants its user. The phrase Open Source emphasizes the right of users to study, change and improve the source code". Most of the free software licensed with General Public License (GNU) is free to modify and share. FOSS, "is a concept and practice of making program source code openly available to everyone to use it" [7].

Usage of software tools and simulators plays important role in today's educational and research to enhance the teaching and learning activities. CAS uses many software and simulation tools for teaching, learning, education management, library management and student information management. Unfortunately most of these tools are proprietary, results in challenges like budget limitation, legal issues in distributing the tools and modifying the source code as the need arise. The wide spread of Internet and evolve of ITC offers great opportunities for learners and changed the way of teaching and learning with considerably low cost, indeed it is essential and unavoidable to provide education to the community at low cost.

This paper intended to identify the list of free and open source tools and software using alternative educational framework [8] for the course of IT programme of CAS. The rest of the paper is organized as follows: section 2 presents the related works, section 3 presents the alternative educational framework, section 4 explains the implementation of strategic areas, section 5 discuss the FOSS suitability for CAS and section 6 provides the conclusion and recommendations.

## II. RELATED WORKS

Several software tools and packages are used in universities and educational institutes for teaching and learning purpose. Many different choices are available in today's market but most of them are expensive [15]. The process of selecting the right tool for academic use is a challenge task due to high cost. In the recent years, the adoption of Free and Open Source Software increased rapidly, countries like Russia make the use of free and open source software mandatory in its entire sectors whereas others are in the process of adopting FOSS. The adoption of FOSS differs from one country to another and from one institution to another however the main reason to this shift is cost reduction. For instance, higher education institutions in United States consider the adoption of Open Source Programs like Moodle more flexible and economical [15]. In addition to cost reduction universities and educational institutes can gain great benefits from adopting FOSS in its courses due to its features which include increasing computing power for students, freedom to make and distribute copies to others without any concerns about software legality, access to source code and security issues [15].

The wide spread of Internet and evolve of information technology offer a great opportunities for learning and changed the way of studying and teaching with a considerably low cost [14]. Universities and many other educational institutes use different modern technologies to distribute knowledge to students; however, most of these technological tools and software are vendor based software. Proprietary software is commercial software, cannot be modified, is licensed and only developed by individual or organization that has the right to develop. Therefore, shifting to Free and Open Source Software will lead to many advantages in teaching and learning environment [13].

Free Software and Open Source Software (OSS) are relatively similar, they are only differing in terms of license mode used, and the use of software after developed for sharing, modification and redistribution [12]. Most of Free Software licensed with General Public License (GNU) which is free to modify and share. Free and Open Source Software "is a concept and practice of making program source code openly available to everyone to use it". Programmers have the chance to open the source code and modifies like adding some new features to suite their local needs. In addition, FOSS programs such as GNU/Linux operating system, OpenOffice.Org, Apache HTTP web server, java programming language and PHP web scripting language are completely free which will benefit the educational institutions in many aspects [9,10,11].

### III. ALTERNATIVE EDUCATIONAL FRAMEWORK

Alternative educational framework [8] is referred as the research methodology to identify the FOSS suitability. The strategic areas of the referred framework are: develop FOSS plan, form monitoring committee, identify stakeholders, identify FOSS tools, train aspirants, develop FOSS policies and guidelines, promote awareness on FOSS and evaluate outcome. The policies of the referred framework [8] are: responsible, adoption, technology, develop, implement, training, teaching and awareness. Figure 1 shows the alternative education framework for FOSS adoption [8] with the strategic areas.

In order to identify the FOSS suitability, a FOSS plan has been developed by the researchers, a monitoring committee has been formed; the committee includes FOSS researchers, Coordinator, System Coordinator, educator and students. The roles and the responsibilities of the committee includes: identifying the stakeholders, identifying the FOSS suitability, train the aspirants, developing policies and guidelines, promoting FOSS awareness and evaluating the outcome. Figure 2 show summarized strategic areas with corresponding implementation policies.

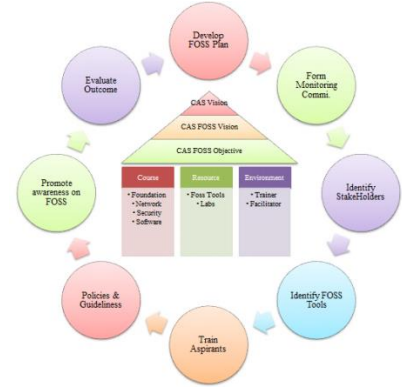

Fig 1: Alternative educational framework for FOSS Adoption

### IV. IMPLEMENTATION OF STRATEGIC AREAS
A. STRATEGIC AREA I

The SAI includes the FOSS technical plan and formation of monitoring committee. The monitoring committee followed a clear line of strategies and supported policies as follows. A FOSS plan was developed with different activities along with clear deadlines and responsibility of each member of the committee by considering the primary objective of migration from the proprietary software to FOSS for all the courses of IT programme. A phased approach was utilized to identify the FOSS tools starts from the foundation year courses to the specialized courses. The monitoring committee ensured the positive movement towards identifying the FOSS suitability and the deviations were noted and reported whenever necessary for a positive movement.

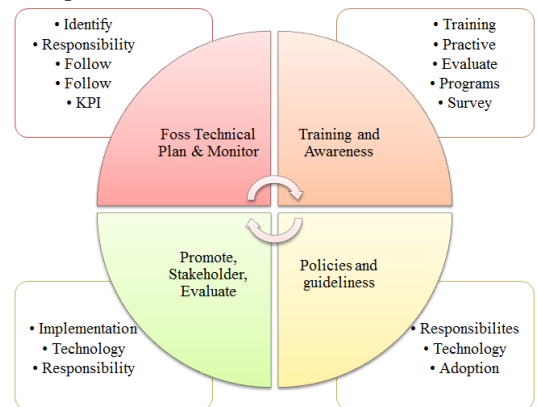

Fig.2: Compressed Strategic areas and policies

B. STRATEGIC AREA II

The SAII includes the development of policies and guidelines by the monitoring committee for adopting the FOSS tools and software. The policy includes: the level of FOSS usage, the security issues and other legal complications need to be addressed. The general guidelines considered and addressed while developing FOSS policy were: determined the FOSS benefits, development of FOSS governance process, specified the roles and responsibilities in support of the FOSS governance, definition to what extent an educator or student can contribute to open source, determined the relationships with the open source community and developed a documentation plan in support of communication and awareness of the CAS's FOSS governance strategy.

### C. STRATEGIC AREA III

The SAIII includes training and awareness of FOSS. A training schedule has been prepared to train the educators, to start implementing FOSS in their courses and to teach the FOSS tools to the students. In case, students and educators are not familiar to use the FOSS tools and face some challenges in FOSS adoption, awareness campaigns has been conducted to increase the FOSS awareness, helped them to move towards FOSS adoption quickly. Awareness campaigns were also conducted through various social media, workshop and classrooms trainings.

### D. STRATEGIC AREA IV

The SAIV includes the FOSS promotion, identifying stakeholders and evaluation of outcome. The monitoring committee, educators and students are responsible for the FOSS promotion. Educators can promote the usage of FOSS tools by using them in their labs works, assignments and projects. Students can promote the usage of FOSS by adopting them in their learning activities and the monitoring committee promotes the FOSS usage by conducting various awareness programs. A set of stakeholders were identified, their technical knowledge and guidelines were utilized in identifying the FOSS suitability of CAS.

### V. FOSS SUITABILITY FOR CAS

As per the vision of the CAS IT programme and the graduates attributes of BIT programme; a course must be intellectually developed and provide advanced knowledge and practical abilities in a range of IT fields [1]. The graduates of CAS must be able to work in a team or unsupervised, must have advanced problem solving skills, pursue lifelong learning and expected to have multidisciplinary and multicultural perspectives. All the graduate attributes [2] of CAS were highly considered and a detailed study has been conducted on various courses of IT programme, considered the alternative educational framework as the reference model. In the FOSS suitability identification process, through the discussion and interview with the educators and based on the outcome from the other stakeholders, the suitable FOSS software for all the courses is listed out. Table 1 show a comparative study between the Blackboard, a proprietary educational management tool and Moodle, an open source educational management tool. A checklist of the features provided by the proprietary software was listed and the same were checked with the features of the open source software. The target of 80 percentage of match was fixed by the monitoring committee; any tool which reaches the target was selected to replace the existing proprietary software.

|  | Blackboard | Moodle |
|---|---|---|
| Grading | √ | √ |
| Assignments | √ | √ |
| Assessments | √ | √ |
| Reports | √ | √ |
| Survey and course evaluation | √ | X |
| Discussion board | √ | √ |
| Mobile support | √ | √ |
| Track progress | √ | √ |
| Personalization | √ | √ |
| Collaborative tools | √ | √ |
| Security | √ | √ |
| Rubrics | √ | √ |
| Communication | √ | √ |

Table 1: Comparison of proprietary & FOSS tool

Table 2 show the list of FOSS software for the individual course of IT programme of CAS, in the plethora of FOSS software, identifying the suitable software for a specific course and its inadequate usage are the common issues exist in adopting them. Listed open source software provides the educators and the learners an ideal opportunity to apply the security concepts in real time and attain the practical experiences. In this research, other than the academic courses, the availability and the suitability of FOSS tools for the education management were also analyzed. Proprietary software are currently in use, for various activities like e-learning, lecture capturing, podcasting, content management, library management, plagiarism checking, classroom management and attendance system. Table 3 show the list of proposed FOSS tools for effective education management in CAS.

### VI. CONCLUSION AND RECOMMENDATIONS

In this paper, several feasibilities to implement FOSS in the IT programme of CAS were analyzed by using alternative educational framework as a reference model, various strategic areas of the reference model; technical plan, monitoring committee, trainers, awareness programme, policies and guidelines, promotions, stakeholders and evaluation are highly considered while identifying FOSS suitability. As a result, a list of alternative FOSS tools and software were identified for the courses of various specializations of IT programme. By adopting the recommended FOSS tools and software, CAS might extremely benefit from the copy right issues, low cost software and availability of source code for learners. The student community of CAS could be highly benefited with the rising demand for the latest technology on the campus and sharing of content may become simple without distressing about software licensing. Adoption of FOSS may also considerably reduce the budget of software purchase, which directly contributes towards nation's economic growth. A committee must be formed for each specializations, comprises system coordinators, programme director, stakeholders and FOSS experts to review and to amend the course outcomes of the IT programme of CAS relevant to the availability of FOSS tools is highly recommended.

| COURSE CODE | COURSE NAME | SOFTWARE IN USE | FOSS SOFTWARE |
|---|---|---|---|
| COMP 4001 / 5001 | Computer Skills A & B | Microsoft office | Apache open office / Libre office / Neo office / KOfficeGIMP Paint.NET / Wildfly-Jboss |
| ITDR 1102 | Introduction to Dynamic Web Development | HTML | Glassfish / Apache Geronimo Apache HTTP Server / Apache tomcat |
| ITDR 1103 | Programming Fundamentals | Java | Spring / Struts / Eclipse |
| ITDR 1105 | Web Development | HTML | Bluefish Editor / Open BEXI / Brackets |
| ITDR 2104 | Programming | Java | Spring / Struts / Eclipse |
| ITDR 2105 | Data Structures (1) | Dr. Java | C, C++ |
| ITDR 2106 | Introduction to Databases | Oracle SQL developer | MySQL / Firebird / MariaDB / PostgresSQL |
| ITDR 3102 | Operating Systems | Ubuntu Packet Tracer OpenSSL, Kali Windows server 2008 / Windows XP / Virtual box/ C, Matlab | Zentyal Linux Server Amahi Home Server Apachi Directory Project Clear OS / CentOS Ubuntu Server & Ubuntu Client Gentoo Linux / GNS3 / Cloonix / CORE / IMUNES / Marionnet / Mininet / Netkit / Psimulator2 / Virtualsquare / VNX and VNUML |
| ITNW 3101 | Introduction to Routing and Switching | | |
| ITNW 4109 | Innovation in Network and Security | | |
| ITNW 3104 | Network Technology | | |
| ITNW 3105 | Network Management | | |
| ITNW 4103 | Internetworking | | |
| ITNW 4104 | Wireless Networking | | |
| ITDR 4100 | Software Project Management | MS Project | JMeter / Linux Desktop Testing / Project |
| ITSY 3104 | Computer Security A | Packet tracer Kali Linux | WireShark / Nmap / OSSEC / Security Onion / Metasploit Framework / OpenSSH Kali / Ostinato |
| ITSY 3105 | Network Security A | Kali, Windows server | |
| ITSY 4104 | Computer Security B | Matlab, Kali | |
| ITSY 4105 | Network Security B | Scapy, Kali, Win Server | |

Table 2 : FOSS suitability for CAS-IT Courses

| EDUCATION MANAGEMENT ACTIVITY | RECOMMENDED FOSS TOOL |
|---|---|
| E-learning | Moodle |
| Lecture Capture/Podcasting | OpenCast Matterhorn |
| Online Lectures/ Webinars / Remote Participation | Openmeetings |
| Interactive Content Creation | Xerte |
| e-book Management | Calibre |
| High-Stakes Assessment | Rogo |
| Interactive Whiteboard Software | OpenSankore |
| Classroom Management | Italc |

Table 3: FOSS suitability for CAS Education Management